\documentclass[%
 reprint,
 amsmath,amssymb,
 aps,
]{revtex4-2}

\usepackage[usenames,dvipsnames]{color}
\usepackage{graphicx}
\usepackage{dcolumn}
\usepackage{bm}
\usepackage{hyperref}
\hypersetup{
    colorlinks=true,
    linkcolor=blue,
    filecolor=magenta,      
    urlcolor=cyan,
    pdftitle={Overleaf Example},
    pdfpagemode=FullScreen,
    }
\begin{document}

\preprint{APS/123-QED}

\title{In Horizon Penetrating Coordinates:\\ Kerr Black Hole Metric Perturbation, Construction and Completion}

\author{Fawzi Aly}
 \email{mabbasal@buffalo.edu}
\author{Dejan Stojkovic}
 \email{ds77@buffalo.edu}
\affiliation{%
HEPCOS, Department of Physics, SUNY at Buffalo
}%

\begin{abstract}

We investigate the Teukolsky equation in horizon-penetrating coordinates to study the behavior of perturbation waves crossing the outer horizon. For this purpose, we use the null ingoing/outgoing Eddington-Finkelstein coordinates. The first derivative of the radial equation is a Fuchsian differential equation with an additional regular singularity to the ones the radial one has. The radial functions satisfy the physical boundary conditions without imposing any regularity conditions. We also observe that the Hertz-Weyl scalar equations preserve their angular and radial signatures in these coordinates. Using the angular equation, we construct the metric perturbation for a circularly orbiting perturber around a black hole in Kerr spacetime in a horizon-penetrating setting. Furthermore, we completed the missing metric pieces due to the mass $M$ and angular momentum $J$ perturbations. We also provide an explicit formula for the metric perturbation as a function of the radial part, its derivative, and the angular part of the solution to the Teukolsky equation. Finally, we discuss the importance of the extra singularity in the radial derivative for the convergence of the metric expansion.

\end{abstract}

\maketitle

\section{Introduction}
Most of the astrophysical black holes are expected to be rotating black holes \cite{Bambi_2020,Kumar_Kumar_Ghosh_2020}. From the phenomenological side, it is thus of utmost importance to describe perturbations of the Kerr metric which can then be used to study Hawking radiation \cite{Iso_Umetsu_Wilczek_2006,Murata_Soda_2006,Jiang_Wu_Cai_2006}, quasi-normal modes \cite{Leaver_1985,Yang_Nichols_Zhang_Zimmerman_Zhang_Chen_2012,Corda_Hendi_Katebi_Schmidt_2013,Berti_Cardoso_Kokkotas_Onozawa_2003}, gravitational waves \cite{Lukes_Gerakopoulos_Harms_Bernuzzi_Nagar_2017,Konoplya_Zhidenko_2016,Gong_Cao_Chen_2021}, and many other related phenomena in a rotating spacetime in the framework of General Relativity (GR). The Kerr spacetime is a stationary, axially symmetric, and asymptotically flat solution to Einstein's field equations in GR that describes the gravitational field around a rotating, uncharged black hole with two horizons in the non-extremal cases \cite{Wald, MTW, matthias_blau, griffiths_podolský_2012,carroll_2003}. It also possesses a hidden symmetry encoded in the Killing-Yano tensor, in addition to the time and azimuthal killing vectors \cite{Frolov_2017,Rudiger_1981,Rudiger_1982}. The metric was originally derived by Roy P. Kerr in 1963 as an extension of the Schwarzschild solution for zero-spin \cite{Kerr_1963,teukolsky_2015, Matt_Visser}.
\vspace{1mm}

The Kerr metric is usually expressed in an oblate spheroidal-like coordinates known as the Boyer-Lindquist (BL) coordinates, which reduce to the Schwarzschild coordinates in the zero-spin limit. It is also worth mentioning that the geodesic equations can be separated in these coordinates as they form an integrable system with four constants of motion: energy, axial angular momentum, mass, and Carter's constant \cite{teukolsky_2015}. Nevertheless, the Boyer-Linguist coordinates are ill-defined at the Kerr black hole's horizons akin to the Schwarzschild coordinates at the Schwarzschild's black hole horizon \cite{teukolsky_2015,Matt_Visser,campanelli_khanna_laguna_pullin_ryan_2001}. Lately, through studying the freely falling observers worldline, Sorge was able to generalize the Lemaître coordinates to the Kerr spacetime which are well defined at these horizons \cite{sorge2021kerr}. However, as we aim to study the massless perturbations here, we find it more convenient to work in a horizon-penetrating coordinate adapted to null geodesics, such as ingoing and outgoing Finkelstein-Eddington (IEF/OEF) coordinates which were constructed for the Kerr case long time before the Lemaître coordinates \cite{Teukolsky1_1973,campanelli_khanna_laguna_pullin_ryan_2001}.
\vspace{1mm}

By the time the Kerr solution was derived, the black hole perturbation theory (BHPT) was already mature and often applied to the spherically symmetric spacetimes such as the Schwarzschild one \cite{Nagar_Rezzolla_2005,Regge_Wheeler_1957,Zerilli_1970,mARTELphdthesis}. It was then natural to extend the investigation to the Kerr spacetime \cite{Teukolsky1_1973}. In 2017, Chen and Stein were able to construct metric perturbations up to the first order for the Near Horizon Extreme Kerr (NHEK) \cite{Bardeen_Horowitz_1999} directly by following an isometry-based approach \cite{Chen_Stein_2017} analogous to the one employed to perturb the Schwarzschild spacetime \cite{mARTELphdthesis}. The metric was expressed in a factorized form and then decoupled thanks to the orthogonality of the symmetry-adapted basis. In 2023, Franchini also managed to decouple the linearized Einstein Field equations after employing spherical harmonics decomposition for a slowly rotating Kerr black hole, up to the second order in spin, in a way similar to the Schwarzschild perturbation scheme \cite{Franchini_2023}. Franchini found a generalized version of both the famous Regge-Wheeler equation \cite{Regge_Wheeler_1957} and of Zerilli equation \cite{Zerilli_1970} which describes the odd and even perturbations modes respectively in the Schwarzschild spacetime. Remarkably, the angular mode mixing resulting from the non-zero spin of the black hole was handled by following the scheme provided in \cite{Kojima_1992} for perturbation of spinning stars up to the first order in spin. 
\vspace{1mm}

Unfortunately, the whole Kerr spacetime hasn't been perturbed in an isometry-based fashion so far, to the best of knowledge of the authors \cite{Franchini_2023,Chen_Stein_2017}. Even in a gauge-dependent and coordinate-dependent settings, it is not clear how to find a symmetry-adapted basis needed to achieve metric perturbation separability, nor how to proceed and find a way to decouple linearized Einstein Field equations of the Kerr spacetime. Nevertheless, the study of gravitational, electromagnetic, and scalar massless perturbations for the Kerr spacetime is typically carried out using the Newman-Penrose (NP) formalism. In this approach, the Weyl tensor $C_{\mu\nu\alpha\beta}$ is projected onto the four null tetrad legs $e^\mu_a$, where the first two are chosen to lie along the repeated null directions of the Weyl tensor. The resulting spinors of the projection are complex and known as the Weyl scalars $\psi_n$ where $n=0,1,2,3,4$. With ten degrees of freedom, they encode all the information in $C_{\mu\nu\alpha\beta}$. As the Weyl tensor coincides with the Riemann tensor $R_{\mu\nu\alpha\beta}$ in vacuum regions, perturbations of these scalars describe perturbations of the spacetime curvature. The projection operation can be applied to the tensors involved in the Einstein field equations to obtain the NP equations in terms of the Weyl scalars \cite{chandrasekhar_2009, Pound_Wardell_2022}. Their first-order perturbation gives rise to coupled equations, but fortunately, they can be decoupled, and the resulting equations admit solutions in a factorized form. Press and Bardeen had applied a perturbation scheme to the Schwarczchild case itself before Teuklosky worked out the Kerr spacetime \cite{Price_1972,Bardeen_Press_1973}. Wald later demonstrated that the majority of information about gravitational perturbations is encoded in $\psi_0$ and $\psi_4$ Weyl scalars \cite{Teukolsky1_1973}. Teukolsky's work resulted in a single master partial differential equation that describes all gravitational, electromagnetic, and scalar field perturbations. The equation is separable in the BL/IEF/OEF and any coordinates related to them with Teukolsky's transformation \cite{Teukolsky1_1973}. After the separation of variables, the Teukolsky PDE reduces to two ODEs: the Radial-Teukloksy and the Angular-Teukloksy equations which both belong to the confluent Heun ODE family \cite{borissov2010exact}. Moreover, the Teuklosky equation reduces the equation, in the zero-spin limit; to the well-known Bardeen-Press equations, the master perturbation equation for the Schwarzschild spacetime obtained using the NP formalism \cite{Glampedakis_Johnson_Kennefick_2017}.
\vspace{1mm}

As the Angular and Radial Teukloksy ODEs are confluent Heun ones, they could be solved through using series expansion in the Hypergeometric function and the coulomb wave function \cite{ronveaux_2007} resultant a three-term recursive relation. Mano, Suzuki, and Takasugi worked out such solutions known as the MST for both the Teukolsky ODEs \cite{Mano_1996, Shuhei_1997} and the Regge-Wheeler ODE, which also a confluent Heun equation akin to the Zerilli ODE \cite{Mano_1996RW}. The MST solution for the Teukolsky equation is practical for computational purposes in the low-frequency limit. Also, Teukolsky ODEs could also be solved using the Confluent Heun functions obtained using Taylor series and Laurent series expansions similarly resultant three-term recursive relation \cite{Fiziev_2009wn, borissov2010exact}. Furthermore in \cite{borissov2010exact}, Fiziev and Borissov managed to employ a special class of the confluent Heun functions known as the Heun polynomial, at which the power series truncated at finite power, through utilizing the Homotopic transformation of the Huen ODEs. They have reported that for scalar $s=\frac{1}{2}$ and electromagnetic $s=1$ perturbation only the separation constant will be constrained in order to obtain those special solutions, allowing for continuous frequency spectrum; however, for gravitational perturbation $s=2$, there will be an extra constrain on the frequencies themselves, hence those polynomial solutions are only valid for particular frequencies of the gravitational waves.
\vspace{1mm}

Moreover, as the Teukolsky master equation preserves its singular structure under transformation to IEF/OEF coordinates\cite{campanelli_khanna_laguna_pullin_ryan_2001}; the Radial-Teukolsky is expected to still belong to the confluent Heun family. Nonetheless, the solution to this boundary value problem shouldn't require imposing regularity at the horizon compared to the case in the BL coordinates \cite{van_de_meent_shah_2015}.
\vspace{1mm}

Yet, it is desirable to obtain the explicit form of the metric perturbation from the curvature perturbation up to a gauge. This problem is known in the literature as the metric reconstruction problem \cite{Pound_Wardell_2022}. For instance, in self-force analysis \cite{Pound_Wardell_2022}, the metric perturbation is needed for further computations. In \cite{lousto_whiting_2002}, Lousto and Whiting investigated the influence of radiation on a particle orbiting a massive rotating black hole. The reconstruction could be achieved in the outgoing/ingoing radiation gauge (ORG/IRG) through a procedure developed by Chrzanowski, Cohen, Kegeles, and Wald known as the CCKW procedure \cite{kegeles_cohen_1979,chrzanowski_1975}. Initially, the CCKW procedure relied on the postulate that the perturbation metric could be obtained using Hertz-like potentials. This postulate was suggested by Chrzanowski and Cohen, and later by Kegeles, for gravitational and electromagnetic perturbations, respectively \cite{kegeles_cohen_1979,chrzanowski_1975}. Wald showed that the success of this technique follows from the adjoint structure of the Teukolsky equation itself \cite{wald_1978}. In \cite{wald_1978}, Wald also showed that not all possible perturbations are contained within the perturbed $\psi_0$ and $\psi_4$. For instance, the mass $M$ and angular momentum $J$ of the black hole itself could be perturbed; those perturbations are not captured by the CCKW. This leads to another problem known as the completion problem \cite{merlin_ori_barack_pound_van_de_meent_2016}, where the missing parts of the full metric perturbation are investigated. Luckily, Wald's theorem proves that four missing parts, up to a gauge, can be constructed from the variations in the mass $M$, angular momentum $J$ , C-metric acceleration $\alpha_C$, or NUT charge $q_{NUT}$ of the black hole. 
\vspace{1mm} 

the CCKW procedure will be well defined for vacuum spacetime; If the perturber is located at a constant radius (e.g. particle in a circular orbit around the Kerr black hole moving), this confines the non-vacuum region to only a hypersurface of $r=const$ in the spacetime \cite{shah_friedman_keidl_2012}. However, if the perturber is also moving radially (e.g. a particle is moving in an elliptic orbit), then the procedure will run into trouble as the ORG/IRG are singular in non-vacuum regions, consequently, CCKW will be ill-defined there. Nevertheless, the metric construction for the elliptical orbits was tackled in \cite{merlin_ori_barack_pound_van_de_meent_2016,van_de_meent_shah_2015}, the authors adopted the method of extending the homogeneous solution to confine the singularity on the trajectory of the perturber relying on the fact that this singularity only exists on the trajectory in the $(t,r)$ hypersurfaces, thus it could be as if there only two hypersurfaces with constant radial coordinates in the frequency domain. Still, generically working in radiation gauges, which is obligated by the CCKW; constrains the matter sources the metric could be constructed for.

Handling arbitrary matter sources was approached in \cite{Green_Hollands_Zimmerman_2020}, through the use of a correction term to make sure that the constructed metric perturbation satisfies the linearized Einstein field equations. Moreover, in \cite{Dolan_2023} a parallel method in Lorenz gauge to the CCKW was recently introduced that overcomes those obstacles brought by the radiation gauges. Moreover, this is not the only way around the CCKW, In \cite{Loutrel_2021}, in ORG the researchers constructed the metric directly without intermediate Hertz-like potentials. Furthermore, taking this as a starting point; they were able to study second-order gravitational perturbation that results in the Teukolsky equation in the second-order perturbation with source term quadratic in the first-order perturbation. 
\vspace{1mm}

There has been recent progress in the construction of the full perturbation metric and in addressing the completion problem in the BL coordinates and Kinnersley tetrads for perturbed orbits \cite{merlin_ori_barack_pound_van_de_meent_2016,van_de_meent_shah_2015,shah_friedman_keidl_2012}. However, this formalism is not suitable to describe perturbations in the near and across-horizon regions. Instead, a null rotation of the tetrads is typically used to impose a regularity condition. The master perturbation equation has also been written in horizon-penetrating coordinates and tetrads, such as the IEF/OEF coordinates \cite{campanelli_khanna_laguna_pullin_ryan_2001}. In these coordinates, the tetrad itself is horizon-penetrating, which eliminates the need for imposing regularity conditions. 
\vspace{1mm}

In this work, we will investigate the leading-order correction to the Kerr metric due to a perturber circularly orbiting the black hole; the analysis will be conducted in a horizon penetrating tetrad and coordinates using the CCKW in radiation gauges for simplicity. Consequently, the full metric is expected to be regular at the horizons as the background metric is written in IEF/OEF and the metric perturbation is constructed in a horizon-penetrating setting. The authors believe this to be vital in performing calculations to study near-horizon phenomena in which the full metric is needed.
\vspace{1mm}

The structure of this paper is as follows. In section II we will go through the underlying mathematical tools needed for the metric construction and metric completion. Also, we will introduce the coordinates we are interested in, as well as the tetrad and NP scalars representation in these coordinates. In Section III, the Teukolsky equation will be introduced, and the separation of the variables will be executed in the new coordinates and tetrads. Then, we will study the radial equation, its radial derivatives, and its boundary conditions. We will solve the inhomogeneous radial equation using the Green's function method. In section IV, we will apply the CCKW procedure starting from the Hertz-Weyl scalars defining equations and then algebrizing the angular equation in the IRG. Finally, in Section V we will construct the metric and add the missing pieces up to some constant coefficient.
\section{\label{sec:level1}preliminary\protect\\}

For our purpose, we will follow the definitions given in \cite{campanelli_khanna_laguna_pullin_ryan_2001,bishop_isaacson_maharaj_winicour_1998} for both the coordinate transformation and tetrads definitions. One may observe that the coordinates in these definitions are slightly different from the usual $U$$(V)$ time coordinates for IEF(OEF) coordinates given by Teukolsky \cite{Teukolsky3_1974}. The reason for that is just convenience, while physics remains the same. We will follow the same conventions as in \cite{campanelli_khanna_laguna_pullin_ryan_2001,bishop_isaacson_maharaj_winicour_1998},  and also adopt the geometric units ${c=G=1}$, along with the metric signature $(+,-,-,-)$.

\subsection{\label{sec:level2}Coordinates}

In our analysis, we will use the BL coordinates $(t,r,\theta,\phi)$, IEF coordinates $(\tilde{t},r,\theta,\tilde{\phi})$, and OEF coordinates $(\hat{t},r,\theta,\hat{\phi})$. We follow \cite{campanelli_khanna_laguna_pullin_ryan_2001} and define the latter two from the first through the following transformation in its differential form:

\begin{equation}\label{radialstar}
dr^*= \frac{r^2+a^2}{\triangle} d r.
\end{equation}

\begin{equation}
\begin{array}{cc}
dr=dr, \\
d\theta=d\theta. \\
\end{array}
\end{equation}

\begin{equation}\label{phiandtimeredefinitionIEF}
\begin{array}{cc}
d\tilde{t}=dt - dr + dr^*, \\
d\tilde{\phi}=d\phi +  \frac{a}{\triangle} d r.
\end{array}
\end{equation}

\begin{equation}\label{phiandtimeredefinitionOEF}
\begin{array}{cc}
d\hat{t}=dt + dr - dr^*, \\
d\hat{\phi}=d\phi - \frac{a}{\triangle} d r.
\end{array}
\end{equation}
Where $\triangle=r^2-2Mr+a^2$.

\subsection{Metric}
Accordingly, the metric in the BL, IEF, and OEF coordinates respectively will take the following structure

\begin{equation} 
  \begin{aligned}
d s^2 & = (1-2 M r / \Sigma) d t^2-(\Sigma / \triangle) d r^2-\Sigma d \theta^2\\
&+\left(4 M a r \sin ^2 \theta / \Sigma\right) d t d \varphi\\
&-\sin ^2 \theta\left(r^2+a^2+2 M a^2 r \sin ^2 \theta / \Sigma\right) d \varphi^2.
  \end{aligned}
\end{equation}

\begin{equation}
\begin{aligned}
d s^2=&(1-2 M r / \Sigma) d \tilde{t}^2-(1+2 M r / \Sigma) d r^2-\Sigma d \theta^2\\
&-\sin ^2 \theta\left(r^2+a^2+2 M a^2 r \sin ^2 \theta / \Sigma\right) d \tilde{\phi}^2 \\
&-(4 M r / \Sigma) d \tilde{t} d r+\left(4 M r a \sin ^2 \theta / \Sigma\right) d \tilde{t} d \tilde{\phi}\\
&+2 a \sin ^2 \theta(1+2 M r / \Sigma) d \tilde{r} d \tilde{\phi}.
\end{aligned}
\end{equation}

\begin{equation}
\begin{aligned}
d s^2=&(1-2 M r / \Sigma) d \hat{t}^2-(1+2 M r / \Sigma) d r^2-\Sigma d \theta^2,\\
&-\sin ^2 \theta\left(r^2+a^2+2 M a^2 r \sin ^2 \theta / \Sigma\right) d \hat{\phi}^2, \\
&+(4 M r / \Sigma) d \hat{t} d r+\left(4 M r a \sin ^2 \theta / \Sigma\right) d \hat{t} d \hat{\phi},\\
&+2 a \sin ^2 \theta(1+2 M r / \Sigma) dr d \hat{\phi}.
\end{aligned}
\end{equation}
Where $\Sigma=r^2+a^2 \cos^2\theta$.
The invariance of the metric under simultaneous $(t,\phi)$ parity is not manifested in the IEF or OEF coordinates. Instead, using the coordinate transformation equations it can be shown that the IEF and OEF coordinates are mapped to one another, though with negative time and azimuthal angle.
\begin{equation}\label{parityIEF}
\begin{array}{cc}
d\hat{t}=-d\tilde{t}, \\
d\hat{\phi}=-d\tilde{\phi}.
\end{array}
\end{equation}

\subsection{Tetrads}
In the case of BL and OEF coordinates, the Kinnersley tetrads will be used. The tetrads will take the following form for each case respectively:

\begin{equation}
\begin{aligned}
l^\mu &=\left[\left(r^2+a^2\right) / \triangle, 1,0, a / \triangle\right], \\
n^\mu &=\left[r^2+a^2,-\triangle, 0, a\right] /(2 \Sigma), \\
m^\mu &=[i a \sin \theta, 0,1, i / \sin \theta] /(\sqrt{2}(r+i a \cos \theta)).
\end{aligned}
\end{equation}

\begin{equation}
\begin{aligned}
l^\mu &=[1,1,0,0], \\
n^\mu &=\left[\frac{\triangle}{2 \Sigma}\left(1+\frac{4 M r}{\triangle}\right),-\frac{\triangle}{2 \Sigma}, 0, \frac{a}{\Sigma}\right], \\
m^\mu &=\left[i a \sin \theta, 0,1, \frac{i}{\sin \theta}\right] /(\sqrt{2}(r+i a \cos \theta)).
\end{aligned}
\end{equation}

When working in the IEF coordinates, we will use the usual Kinnersley tetrads after applying a null rotation of the third kind by rescaling the $l^\mu$  by $\triangle$,  and dividing $n^\mu$  by a factor of $\triangle$. Using these modified Kinnersley tetrads we expect that epsilon, which was set to zero in Teukolsky's paper \cite{Teukolsky1_1973} by the null rotation freedom, will be in general non-zero. In these coordinates, the tetrads will take the following form

\begin{equation}
\begin{aligned}
l^\mu &=[\triangle+4 M r, \triangle, 0,2 a], \\
n^\mu &=\left[\frac{1}{2 \Sigma},-\frac{1}{2 \Sigma}, 0,0\right],\\
m^\mu &=\left[i a \sin \theta, 0,1, \frac{i}{\sin \theta}\right] /(\sqrt{2}(r+i a \cos \theta)).
\end{aligned}
\end{equation}

\subsection{NP Scalars}
The non-vanishing NP scalars will take the same form in all coordinates:
\begin{equation}
\begin{aligned}
\beta &=\frac{\cot \theta}{2 \sqrt{2}(r+i a \cos \theta) },  \\
\pi &=\frac{i a \sin \theta }{\sqrt{2}(r-i a \cos \theta)^2},  \\
\tau &=\frac{-i a \sin \theta}{\sqrt{2} \Sigma},  \\
\alpha &=\pi-\beta^*, \\
\psi_2 &=\frac{-M}{(r-i a \cos \theta)^3}.
\end{aligned}
\end{equation}
Where the complex conjugate is indicated by $*$ symbol. While in BL and OEF coordinates, the rest of the non-vanishing NP quantities will take this form 

\begin{equation}
\begin{aligned}
\rho &=\frac{-1}{r-i a \cos \theta}, \\
\mu &=\frac{\triangle}{\Sigma} \frac{-1}{2(r-i a \cos \theta)}, \\
\gamma &=\mu+\frac{r-M}{2 \Sigma}. \\
\end{aligned}
\end{equation}
On the other hand, in IEF coordinates, the rest of the non-vanishing NP quantities will be
\begin{equation}
\begin{aligned}
&\epsilon=r-M, \\
&\gamma=\mu=-\frac{1}{2} \frac{r+i a \cos \theta}{\Sigma^2}, \\
&\rho=-(r+i a \cos \theta) \frac{\triangle}{\Sigma}.
\end{aligned}
\end{equation}
 We can see that the coordinates, tetrads, and NP quantities in IEF and OEF coordinates are well-behaved at the horizon, except for $\rho$ in IEF and $\mu$ in OEF coordinates. 

\section{Solving Teukolsky Master Equation}
\subsection{Teukolsky Master Equation}

As mentioned in the introduction, Teukolsky equations are derived within the NP formalism which makes it coordinate invariant. Once a set of coordinates and tetrads are chosen, the NP equations are expressed in those coordinates as well as the tetrad in use. The two Teukolsky equations of interest for us are the ones defining $\psi_0$ and $\psi_4$. They have the following form in the NP formalism \cite{keidl_friedman_wiseman_2007}.

\begin{equation}
\begin{gathered}
{\left[\left(D-3 \epsilon+\epsilon^*-4 \rho-\rho^*\right)(\Delta-4 \gamma+\mu)\right.} \\
\left.-\left(\delta+\pi^*-\alpha^*-3 \beta-4 \tau\right)\left(\delta^*+\pi-4 \alpha\right)-3 \psi_2\right] \psi_0, \\
=4 \pi T_0,
\end{gathered}
\end{equation}

\begin{equation}\label{FourthEquation}
\begin{gathered}
{\left[\left(\Delta+3 \gamma-\gamma^*+4 \mu+\mu^*\right)(D+4 \epsilon-\rho)\right.} \\
\left.-\left(\delta^*-\tau^*+\beta^*+3 \alpha+4 \pi\right)(\delta-\tau+4 \beta)-3 \psi_2\right] \psi_4, \\
=4 \pi T_4,
\end{gathered}
\end{equation}
where the four-tetrad derivative is: 
\begin{equation}
  \begin{aligned}
    D & =l^\mu \partial_\mu, \\
    \Delta & =n^\mu \partial_\mu, \\
    \delta&=m^\mu \partial_\mu,  \\
     \delta^*&={m^*}^\mu \partial_\mu. \\
  \end{aligned}
\end{equation}
We consider now the master Teukolsky equation for $|s|=2$. The Teukolsky equation in the BL coordinates is defined as  \cite{Teukolsky1_1973}
\begin{equation}
\begin{gathered}
\left\{\left[\frac{\left(r^2+a^2\right)^2}{\Delta}-a^2 \sin ^2 \theta\right] \frac{\partial^2}{\partial t^2}-2 s\left[\frac{M\left(r^2-a^2\right)}{\Delta}\right.\right. \\
-r-i a \cos \theta] \frac{\partial}{\partial t}+\frac{4 M a r}{\Delta} \frac{\partial^2}{\partial t \partial \phi} \\
-\Delta^{-s} \frac{\partial}{\partial r}\left(\Delta^{s+1} \frac{\partial}{\partial r}\right)-\frac{1}{\sin \theta} \frac{\partial}{\partial \theta}\left(\sin \theta \frac{\partial}{\partial \theta}\right) \\
-2 s\left[\frac{a(r-M)}{\Delta}+\frac{i \cos \theta}{\sin ^2 \theta}\right] \frac{\partial}{\partial \phi} \\
\left.+\left[\frac{a^2}{\Delta}-\frac{1}{\sin ^2 \theta}\right] \frac{\partial^2}{\partial \phi^2}+\left(s^2 \cot ^2 \theta-s\right)\right\} {}_s \psi_{lm} \\
=4 \pi \Sigma T_s.
\end{gathered}
\end{equation}

In IEF and OEF coordinates, the same equation will take the following form \cite{campanelli_khanna_laguna_pullin_ryan_2001}:

\begin{equation}
\begin{gathered}
\left\{[\Sigma+2 M r] \frac{\partial^2}{\partial t^2} \pm[2(s \mp 1) M\right. \\
+2 s(r+i a \cos \theta)] \frac{\partial}{\partial t} \mp 4 M r \frac{\partial^2}{\partial t \partial r} \mp 2 a \frac{\partial^2}{\partial t \partial \phi} \\
-\triangle^s \frac{\partial}{\partial r}\left(\triangle^{-s+1} \frac{\partial}{\partial r}\right)-\frac{1}{\sin \theta} \frac{\partial}{\partial \theta}\left(\sin \theta \frac{\partial}{\partial \theta}\right) \\
\mp 2 s\left(\frac{i \cos \theta}{\sin ^2 \theta}\right) \frac{\partial}{\partial \phi} \\
\left.-\frac{1}{\sin ^2 \theta} \frac{\partial^2}{\partial \phi^2}+\left(s^2 \cot ^2 \theta \pm s\right)\right\} {}_s \psi_{lm}\\
=4 \pi \Sigma T_s^{\pm}.
\end{gathered}
\end{equation}
where upper refers to IEF while lower refers to OEF coordinates. The quantity ${}_s \psi_{lm}$ is defined as:
\begin{equation}
{}_s \psi_{lm}= \begin{cases}(r-ia\cos \theta)^4 \psi_4 \quad   s=-2, \\
\psi_0  \qquad \qquad \qquad \quad  s=+2.\end{cases}
\end{equation}
From equations (\ref{parityIEF}) and (\ref{FourthEquation}), it could be easily shown that the $\psi_0$ of IEF coordinates gets mapped to $\psi_4$ of OEF coordinates and vice versa.
\subsection{Separation}
In \cite{Teukolsky3_1974} and \cite{Teukolsky1_1973}, Teukolsky conjectured that the separability should be granted under the transformation between coordinates which takes the following form
\begin{equation}
\begin{array}{cc}
 &\bar{r}=h(r),\\
 &\bar{\theta}=j(\theta),\\
&\bar{t}=t+f_1(r)+f_2(\theta),\\
&\bar{\varphi}=\varphi+g_1(r)+g_2(\theta).\\
\end{array}
\end{equation}
Using Fourier decomposition for time and azimuthal angle, the master equation can be separated in the IEF and OEF coordinates with the following ansatz 
\begin{equation}
{}_s \psi_{lm}=e^{-i \omega t} e^{i m \phi} { }_s S_{l m}(\theta) { }_s R_{l m}(r).
\end{equation}
This will yield the same angular equation as the one in BL coordinates
\begin{equation}
\begin{aligned}
& \left\{\frac{1}{\sin \theta} \frac{d}{d \theta}\left(\sin \theta \frac{d}{d \theta}\right)-\frac{(m+s \cos \theta)^2}{\sin ^2 \theta}-2 a \omega s \cos \theta\right. \\
& \left.+a^2 \omega^2 \cos ^2 \theta+S+{ }_s A_{lm}\right\}{ }_s S_{lm}(\theta)=0.
\end{aligned}
\end{equation}
\vspace{.1cm}
The radial equation has the following forms in IEF and OEF coordinates respectively:
\begin{equation}
\begin{aligned}
 &\triangle \left \{\frac{[-\bar{\lambda}+\omega^2(\triangle+4 M r)+2 i M(s-1) \omega+2 i r s \omega-2 s]}{\triangle}\right. \\
&\left.+\frac{2(i a m+(1-s)(r-M)-2 i M r \omega)}{\triangle} \frac{d}{d r}+  \frac{d^2}{d r^2} \right \}{}_s R(r)_{l m}\\
&\qquad \qquad \qquad \qquad\qquad ={}_sT^{+}_{lm},
\end{aligned}
\end{equation}

\begin{equation}
\begin{aligned}
 &\triangle \left \{\frac{[-\bar{\lambda}+\omega^2(\triangle+4 M r)+2 i M(s+1) \omega+2 i r s \omega]}{\triangle}\right. \\
&\left.+\frac{2(i a m-(1+s)(r-M)-2 i M r \omega)}{\triangle} \frac{d}{d r}+  \frac{d^2}{d r^2} \right \}{}_s R(r)_{l m}\\
&\qquad \qquad \qquad \qquad\qquad ={}_sT^{-}_{lm},
\end{aligned}
\end{equation}
\vspace{.1 cm}
where $\lambda$ and $\bar{\lambda}$  are defined below
\vspace{.1 cm}

\begin{equation}
\begin{array}{cc}
\lambda \equiv A+a^2 \omega^2-2 a m \omega,\\
\bar \lambda \equiv \lambda+2 am\omega.\\
\end{array}
\end{equation}
The radial source term is defined by expanding the source terms for the corresponding Teukolsky equation in the spin-weighted angular harmonics.

\subsection{Radial Equation}

It can be easily seen that the radial equation of the Teukolsky master equation has 3 singular points at $r=\{r_{-},r_{+},r \rightarrow \infty\}$ of rank $\{1,1,2\}$ respectively. The first two singularities at the horizons are regular, while the one at infinity is irregular. Furthermore, the following transformation will put the radial equation written in IEF/OEF coordinates in the regular form of the confluent Heun differential equation (CHE)
\begin{equation}
   R_{\pm}=\hat{R}_\pm e^{\mp i \omega r}.
\end{equation}
This is not surprising, as the transformation from the BL coordinates to the Kerr-ingoing/outgoing coordinates does not alter the radial coordinate at the same time the reader can check that also the redefinition of the azimuthal and temporal coordinates won't change the singular structure of the radial equation by applying chain rule. It should be noted that all confluent Heun ODEs are interrelated through a radial coordinate transformation, as outlined in \cite{ronveaux_2007}. In other words, there exists a radial coordinate transformation that is equivalent to the transformations performed in the azimuthal and temporal coordinates as given by equations (\ref{phiandtimeredefinitionIEF}) and (\ref{phiandtimeredefinitionOEF}).
\vspace{.1 cm}

The coefficient for the second derivative of the dependent variable and the coefficient for the dependent variable itself are second-degree polynomials in $r$, while the coefficient for the first derivative of the dependent variable is a first-degree polynomial. 
\begin{equation}
P(r) \hat{R}_{+}^{\prime \prime}(r)+\tilde{P}(r) \hat{R}_{+}^{\prime}(r)+ \bar{P}(r) \hat{R}_{+}(r)=0,
\end{equation}
Where
\begin{equation}
\begin{gathered}
\bar{P}(r)=-\kappa-2 s+2 i r\omega(2 s-1), \\
\tilde{P}(r)=-2 i a^2 \omega+2 i a m+2 M(s-1)+2r(1-s)-2 i \omega r^2, \\
P(r)=\triangle(r).\\
\end{gathered}
\end{equation}
Similarly, we will have 
\begin{equation}\label{ploynomialFormsinOEF}
Q(r) \hat{R}_{-}^{\prime \prime}(r)+\tilde{Q}(r) \hat{R}_{-}^{\prime}(r)+ \bar{Q}(r) \hat{R}_{-}(r)=0,
\end{equation}
Where
\begin{equation}
\begin{gathered}
\bar{Q}(r)=-\kappa+2i\omega r(2s+1),\\
\tilde{Q}(r)=2 i a^2 \omega-2 i a m-2 M(s+1)+2 r(s+1)+2i\omega r^2,\\
Q(r)=\triangle(r).\\
\end{gathered}
\end{equation}
where all of $Q,\tilde{Q},\bar{Q},P,\tilde{P},\bar{P}$ are polynomials of $r$. Further investigation is needed to solve these equations using the series methods within the region $r_{-} < r < r \rightarrow \infty$ for a better understanding of the behavior of gravitational perturbations after crossing the outer horizon. 

\subsection{First Derivative of the Radial Function}
We need to comprehend the behavior of the first derivative of the radial function to study the perturbed metric expansion. For this purpose, it would be more convenient to put the CHE in the canonical form \cite{ronveaux_2007}. For simplicity, we will only tackle the radial equation in the IEF coordinates in this section
\begin{equation}
\hat{R}^{\prime \prime}(r)+\hat{R}^{\prime}(r)\left(\frac{\alpha}{r-1}+\frac{\gamma}{r}+\epsilon\right)+\frac{(\xi \\ r-\beta)}{(r-1) r} \hat{R}(r)=0.
\end{equation}
Here, $r \rightarrow \dfrac{r-r_{-}}{r_{+}-r_{-}}$. We will rename the independent variable to $r$ in these coordinates, hoping that it will not be a source of confusion. The parameters in the equation above are defined as follows:
\begin{equation}
\begin{aligned}
\epsilon&=4 i \sqrt{-a^2+M^2} \omega, \\
\gamma&=1-s+2 i M\omega-\frac{i\left(a m-2 M^2 \omega\right)}{\sqrt{-a^2+M^2}}, \\
\alpha&=1-s+2 i M\omega+\frac{i\left(a m-2 M^2 \omega\right)}{\sqrt{-a^2+M^2}}, \\
\xi&=4 \sqrt{-a^2+M^2} \omega(-i+4M\omega), \\
\beta&=\omega\left(2 am+2 iM(-1+s)+a^2 \omega\right)-2M\omega(-i+4M\omega),\\
&+2 \sqrt{-a^2+M^2} \omega(-i+4M\omega )+\kappa.
\end{aligned}
\end{equation}
After some algebra, the equation governing the first derivative of the radial function labeled $u(r)$ below can be written as 
\begin{equation}
\begin{aligned}
&u^{\prime \prime}(r)+u^{\prime}(r)\left(\frac{\alpha+1}{r-1}-\frac{1}{\dfrac{\beta}{\xi}- r}+\frac{\gamma+1}{r}+\epsilon\right)\\
&+u(r)\left(\frac{A}{r-1}+\frac{B}{r-\dfrac{\beta}{\xi}}+\frac{C}{r}\right)=0.
\end{aligned}
\end{equation}
\begin{equation}
\begin{aligned}
A &=\gamma+\epsilon+\beta\left(-1+\frac{\alpha}{\beta-\xi}\right)+\xi, \\
B &=\alpha-\epsilon-\frac{\gamma \xi}{\beta}+\frac{\alpha \beta}{-\beta+\xi}, \\
C &=-\alpha+\beta-\gamma+\epsilon+\frac{\gamma \xi}{\beta}.\\
\end{aligned}
\end{equation}

In \cite{filipuk_ishkhanyan_dereziński_2020}, it was shown that this equation has one additional regular singularity point at $r=\frac{\beta}{\xi}$. If this point didn't coincide with $\{0,1,\infty\}$ then the radius of convergence of any function written in terms of the radial equation and its derivative will be the intersection between their two radii of convergences. Thus, we need to account for this in the metric expansion if needed.
\vspace{1mm}

This fact will be crucial once we try to expand the metric perturbation in the radial solution and its derivative, as we will discuss later. Moreover, this new ODE has the irregular singularity of rank $2$ at infinity beside three regular singular points at $r=\left\{0,1,\dfrac{\beta}{\xi}\right\}$. Hence it doesn't belong to the Heun family which has a similar singular structure but has also regular singularity at infinity.

\subsection{Black hole boundary conditions}

We will use Green's functions method to write the inhomogeneous radial solution in terms of the homogeneous radial solutions obeying the boundary conditions imposed on the radial part. There are two physical boundary conditions that must be satisfied at the Horizon $r=r_{+}$ and in the asymptotically flat region $r \rightarrow \infty$. For example, an observer near the Horizon should not see anything special occurring at Horizon. This requires the coordinates, tetrads, and radial part of the Weyl scalars to be regular there. Similarly, an observer at infinity should expect to receive a spherical wave with the same frequency as the frequency of the perturber. We study the homogeneous radial equation  which asymptotically has wave solutions by following \cite{Teukolsky1_1973}. We transform the radial equation to a general harmonic oscillator equation by transforming the dependent and independent variables before studying any limits. The independent variables are given by equation (\ref{radialstar}), while the dependent ones are given below. We include the subscript $s$ since the functions $f$ will be $s$ dependent.
\begin{equation}
{}_{s} R(r)_{lm}= Y(r) f(r),
\end{equation}
where the defining equation for $f(r)$ is 
\begin{equation}
f_{\pm, r *}+\eta_{\pm}f_\pm=0.
\end{equation}
Using those definitions make the equations of $Y(r)$ as follows
\begin{equation}\label{GeneralAsymoticEq}
Y_{, r * r *}+\left\{\frac{\beta_{\pm} \triangle}{(\triangle-2 M r)^2}-\eta_{\pm}^2-\frac{\triangle \eta_{\pm}^{\prime}}{\triangle-2 M r}\right\} Y=0,
\end{equation}
While $\eta_\pm$ is defined as
\begin{equation}
\begin{aligned}
\eta_{\pm}(r) &\equiv \frac{M\left(a^2-r^2\right)}{\left(a^2+r^2\right)^2}\\
&+\frac{\pm i a m+(\mp s+1)(r-M) \mp 2 i M r w}{a^2+r^2}.
\end{aligned}
\end{equation}
which will allow us to solve for $f(r)$
\begin{equation}
f_\pm(r)\equiv \frac{\triangle ^{\pm \left(\frac{s}{2}+i M w\right)} e^{\pm i \alpha  \tanh ^{-1}\left(\frac{r-M}{r_{+}-M}\right)}}{\sqrt{\triangle +2
   M r}},
\end{equation}
where $\alpha$ is defined as
\begin{equation}
\alpha \equiv \frac{a m-2 M^2 w}{r_{+}-M}.
\end{equation}

\subsubsection{At the outer horizon $r \rightarrow r_+$\\"Blackhole"}

As the $r \rightarrow r_+$ then $\triangle \rightarrow 0$. Given that $\dfrac{dr}{dr*} \rightarrow 0$, then $\eta_\pm$ can be treated as a constant with respect to $r*$. Then equation (\ref{ploynomialFormsinOEF}) with its solution will take the following form
\begin{equation}
\begin{gathered}
Y_{, r * r *}(r * \rightarrow -\infty)-\eta^2_\pm Y(r * \rightarrow -\infty) \approx 0,  \\  
Y_{, r * r *}(r * \rightarrow -\infty) \approx  e^{\pm \eta_\pm  r *}.
\end{gathered}
\end{equation}
A similar argument could be applied to the $f(r)$ defining equation, which will leave us with the following solution for $f(r)$
\begin{equation}
f(r* \rightarrow -\infty ) \approx  e^{- \eta_\pm  r *}.
\end{equation}
Finally, $R(r)$ can be evaluated at the Horizon
\begin{equation}
R(r) \approx
\left\{
    \begin{array}{lr}
      1 \text{\hspace{.88cm} (+) ingoing,\hspace{.3cm}  (-) outgoing},   \\
      e^{-2 \eta_\pm  r *} \text{(+) outgoing,\hspace{.10cm} (-) ingoing}   
    \end{array}
\right\} 
\end{equation}
To examine what this means, it would be useful to rewrite $\eta_\pm$ as
\begin{equation}
\eta_\pm (r) = \frac{r \triangle}{(\triangle+2Mr)^2}\mp \frac{i\omega (2Mr) -iam}{\triangle+2Mr}\mp \frac{s(r-M)}{\triangle+2Mr}.
\end{equation}
Thus, at $r \rightarrow r_+$
\begin{equation}
\eta_\pm (r_+) = \mp(i\omega - \frac{iam}{2Mr_+})\mp \frac{s(r_{+}-M)}{2Mr_+}.
\end{equation}
\begin{equation}\label{IEFHorizonBoundary}
R(r) \approx
\left\{
    \begin{array}{lr}
      1 \text{\hspace{1.66cm} (+) ingoing,\hspace{.3cm}  (-) outgoing},   \\
      e^{\pm 2 k r*} \triangle^{\mp s/2} \text{(+) outgoing,\hspace{.10cm} (-) ingoing}   
    \end{array}
\right\}. 
\end{equation}
In the case of the IEF coordinates, $\eta_+$ is proportional to $-i \omega$. Thus, from equation (\ref{IEFHorizonBoundary}) we see that only the ingoing solution is well behaved at the horizon in these coordinates, while in the case of the OEF coordinates, $\eta_-$ is proportional to $i \omega$, so only the outgoing solution is well behaved at the horizon.

\subsubsection{At Infinity $r \rightarrow \infty$\\ }

At infinity, if we expand equation (\ref{GeneralAsymoticEq}) to first order in $1/r$,  we will obtain the asymptotic behavior in \cite{Teukolsky1_1973} for $r \rightarrow\infty$.
\begin{equation}
\begin{gathered}
Y_{, r * r *}(r \rightarrow \infty)+\left(\omega^2+\frac{2 i \omega s}{r}\right) Y(r \rightarrow \infty)\approx 0,\\
Y_{, r * r *}(r * \rightarrow \infty) \approx  r^{\mp s} e^{\pm i\omega r*}.
\end{gathered}
\end{equation}
Now we can evaluate the asymptotic behavior of  $f_\pm (r* \rightarrow \infty)$
\begin{equation}
f_\pm(r \rightarrow \infty)= r ^{-1 \pm \left(s+2i M w\right)} e^{\pm i \alpha  \tanh ^{-1}\left(\frac{r-M}{r_{+}-M}\right)}.
\end{equation}
In the IEF coordinates,
\begin{equation}
R_{+}(r*\rightarrow \infty) =  \frac{r^{+s}}{r^{1 \pm s}} e^{\pm i\omega r*} e^{-i\pi} e^{\pm 2iM\omega  \ln r}.
\end{equation}
In the OEF coordinates,
\begin{equation}
R_{-}(r*\rightarrow \infty) =  \frac{r^{-s}}{r^{1 \pm s}} e^{\pm i\omega r*} e^{-i\pi} e^{\pm 2iM\omega  \ln r}.
\end{equation}

\subsection{Inhomogeneous Radial Equation}

When we attempt to construct the metric in the Outgoing radiation gauge, we will only study the inhomogeneous radial equation for $\psi_4$ for the reasons that will be obvious in the next section. The source term on the right-hand side of the Teukolsky equation for $\psi_4$ is given by the following equation
\begin{equation}
\begin{aligned}
T_{-2} &=8\pi\Sigma S^{\mu \nu}_{-2} T_{\mu \nu}.
\end{aligned}
\end{equation}
$T_{ab}$, the energy-momentum tensor in the tetrad basis will be given as
\begin{equation}
T_{ab}=T_{\alpha \beta} e_a^\alpha e_b^\beta.
\end{equation}
The decoupling operator for the linearized Einstein field equations for $\psi_4$ as provided in \cite{wald_1973} is
\begin{equation}
\begin{aligned}
S^{\alpha \beta}_{-2}=&\{(\left.\Delta+3 \gamma-\gamma^*+4 \mu+\mu^*\right)[\left(\delta^*-2 \tau^*+2 \alpha^*\right) e_4^\alpha e_4^\beta \\
-&\left(\Delta+2 \gamma-2 \gamma^*+\mu^*\right) e_2^\alpha e_4^\beta]\\
+&\left(\delta^*-\tau^*+\beta^*+3 \alpha+4 \pi\right)[\left(\Delta+2 \gamma+2 \mu^*\right) e_2^\alpha e_4^\beta\\
-&\left(\delta^*-\tau^*+2 \beta^*+2 \alpha\right) e_2^\alpha e_2^\beta]\}.
\end{aligned}
\end{equation}

$\mathcal{O}_{-2}$ is the second-order linear radial differential operator representing the radial equation in both the IEF and OEF coordinates respectively. A Green's function can be defined for the following operators (in a way similar to \cite{shah_friedman_keidl_2012})
\begin{equation}
{}_{\pm} \mathcal{O}_{+2}(r) G_\pm\left(r, r^{\prime}\right)=\delta\left(r-r^{\prime}\right),
\end{equation}
Then $G_\pm\left(r, r^{\prime}\right)$ can be written using the homogeneous solution:
\begin{equation}
G_{\pm lm}\left(r, r^{\prime}\right)=\left\{\begin{array}{lr}
c^{\pm}_{lm}(r^{\prime})R^{-}_\pm(r \hspace{.07cm} ) R^{+}_\pm(r^\prime) \quad r_{+}<r<r, \quad\\
c^{\pm}_{lm}(r^{\prime})R^{-}_\pm(r^{\prime}) R^{+}_\pm(r\hspace{.07cm}) \hspace{.4cm} r<r<\infty
    \end{array}
\right\}.
\end{equation}
The superscripts $\{+,-\}$ in $R(r)$ indicate that this quantity satisfies the boundary conditions at infinity and the outer horizon respectively. The coefficient $c^{\pm}_{lm}$  is defined below, while $W[R^{+}_{\pm}(r^\prime),R^{-}_{\pm}(r^\prime)]$ is the Wronskian of the radial equation.
\begin{equation}
c^{\pm}_{lm}(r^{\prime})=\frac{1}{\triangle(r^{\prime})  W[R^{+}_\pm(r^\prime),R^{-}_\pm(r^\prime)]}.
\end{equation}
As shown  in \cite{shah_friedman_keidl_2012}, the full Green's function could still be generated using the completeness of the spin-weight spheroidal harmonics. 

\begin{equation}
\boldsymbol{G\left(x, x^{\prime}\right)}=\sum_{l m} G_{\pm l m}\left(r, r^{\prime}\right) S_{\ell m}(\theta)_2 S_{\ell m}\left(\theta^{\prime}\right) e^{i m\left(\phi-\phi^{\prime}\right)}.
\end{equation}

Finally, we can write $\psi_0$ using Green's function as 
\begin{equation}
\begin{aligned}
\psi_4 &=\int \boldsymbol{G\left(x, x^{\prime}\right)} [8 \pi \Sigma^{\prime} T_{+2}\left(x^{\prime}\right)]  d^3 \vec{r^\prime}. \\
\end{aligned}
\end{equation}
Now we can use the adjoint operator of $S^{\mu \nu}_{+2}$ to simplify this expression
\begin{equation}\label{60}
[\boldsymbol{G\left(x, x^{\prime}\right)} \Sigma] S^{\mu \nu}_{-2} T_{\mu \nu}= T_{\mu \nu} S^{\mu \nu \dagger}_{-2},[\boldsymbol{G\left(x, x^{\prime}\right)} \Sigma] + \partial^i k_i.
\end{equation}
At this point, we can utilize the fact that the energy-momentum tensor of a point particle will always be written as a tensor multiplied by a Dirac delta function. As we are interested in a perturber moving in an equatorial circular orbit around the Kerr blackhole then, the energy-momentum tensor in the coordinate basis is 
\begin{equation}\label{stressEnergyTensor}
\begin{gathered}
T^{\mu \nu} =\frac{E}{\gamma^2} U^\mu U^\nu \delta^3(\vec{r}-\vec{r_p(t)})\equiv \mathcal{F}^{\mu \nu} \delta^3(\vec{r}-\vec{r_p(t)}),\\
\delta^3(\vec{r}-\vec{r_p(t)})=\frac{1}{R^2} \delta(r-R) \delta(\cos(\theta)) \delta(\phi-\Omega t).
\end{gathered}
\end{equation}
where $\Omega$ is the angular frequency of the particle, while $\vec{r_p}$ represents the position vector of the perturber. Then the expression given for $\psi_4$ is
\begin{equation}
\begin{aligned}
\psi_4 =&\int \mathcal{F}^{\mu \nu}(x^{\prime \mu}) \delta^3(\vec{r^\prime}-\vec{r_p(t)}) S^{\mu \nu \dagger}_{+2} [\boldsymbol{G\left(x, x^{\prime}\right)} \Sigma] d^3 \vec{r^\prime}\\
+& \int \partial^i k_i d^3 \vec{r^\prime}.\\
\end{aligned}
\end{equation}
Since $S^{\mu \nu \dagger}_{-2}$ is a second-order linear differential operator, $k^i$ is a function of the particle Dirac delta and its derivatives. Thus, the contribution from the second integral will be zero.
Furthermore, as $\boldsymbol{G\left(x, x^{\prime}\right)}$ is a factorized function in $x$ and $x^{\prime}$, we can write 
\begin{equation}
\begin{gathered}
\boldsymbol{G(x, x^{\prime}})=\sum_{lm} \mathcal{G}_{lm}(x) \mathcal{\tilde{G}}_{lm}(x^\prime),\\
\mathcal{G}_{lm}(x)= {}_2 \mathcal{R}_{lm}(r)    {}_2 S_{lm}(\theta)  e^{i m\phi},\\
\mathcal{\tilde{G}}_{lm}(x^\prime)= {}_2 \mathcal{\tilde{R}}_{lm}(r^\prime)    {}_2 S_{lm}(\theta^\prime)  e^{i m\phi^\prime},
\end{gathered}
\end{equation}
where $\mathcal{R}_{lm}(r)$ and $\mathcal{\tilde{R}}_{lm}(r^\prime) $ is defined as 
\begin{equation}
\begin{aligned}
\mathcal{R}_{lm}(r)=&\left\{\begin{array}{lr}
R^{-}_\pm(r) \quad r_{+}<r<r \quad\\
 R^{+}_\pm(r) \hspace{.4cm} r<r<\infty
    \end{array},
\right\}\\
\mathcal{\tilde{R}}_{lm}(r^\prime)=& \hspace{0.1cm} c^{\pm}_{lm}(r^{\prime}) \left\{\begin{array}{lr}
 R^{+}_\pm(r^\prime) \quad r_{+}<r<r \quad\\
R^{-}_\pm(r^{\prime}) \hspace{.4cm} r<r<\infty
    \end{array}
\right\}.
\end{aligned}
\end{equation}
Since we are in the Fourier space for $(t,\phi)$, then $S^{\mu \nu \dagger}_{-2}$ has no derivatives in both of these coordinates (replaced by their eigenvalues $(\omega,m)$ respectively. Then we can safely get part of  $\mathcal{\tilde{G}}_{lm}(x^\prime)$ which dependents of $\phi$ after taking into account the action of the $\delta(\phi-\Omega t)$ on it. Finally, $\psi_0$ can be written as 
\begin{equation}\label{65}
\begin{gathered}
\psi_4 = \sum_{lm} \mathcal{C}_{lm} \hspace{.05cm} {}_{-2} \mathcal{R}_{lm}(r) \hspace{.05cm} {}_{-2} S_{lm}(\theta)  e^{i m\phi - i\Omega t},\\ 
\mathcal{C}_{lm}=\{\mathcal{F}^{\mu \nu} S^{\mu \nu \dagger}_{-2} [ {}_{-2} \mathcal{\tilde{R}}_{lm}(r^\prime)  \Sigma(r^\prime,\theta^\prime) {}_{-2} S_{lm}(\theta^\prime)]\}_{\vec{r^\prime}=\vec{r}_p(t)}.\\
\end{gathered}
\end{equation}
The adjoint operator $S^{\mu \nu \dagger}_{-2}$ is given as
\begin{equation}
\begin{aligned}
S^{\mu \nu \dagger}_{-2}=& e_4^\alpha e_4^\beta (\delta^*+\tau^*-3 \alpha+\beta^*+\pi)(\Delta-4 \gamma-3 \mu) \\
-& e_2^\alpha e_4^\beta[(\Delta+\mu-3 \gamma+\gamma^*)(\Delta-4 \gamma-3 \mu) \\
-&\left(\Delta+2 \mu-2 \mu^*-3 \gamma-\gamma^*\right)\left(\delta^*+\pi-4 \alpha-4 \tau\right)] \\
-&e_2^\alpha e_2^\beta\left(\delta^*+\pi-3 \alpha-\beta^*\right)\left(\delta^*+\pi-4 \alpha-4 \tau\right).
\end{aligned}
\end{equation}

\section{Applying the CCKW Procedure}

The CCKW procedure is dedicated to constructing the metric from the Hertz potential. To arrive at this final goal, it will be crucial to algebrize the equation connecting the Weyl scalars. Thus, the source terms in the Teukolsky equation would be manifested in the perturbation metric \cite{pound_merlin_barack_2014}. 

\subsection{Hertz Potential-Weyl Scalars Equations}

In CCKW, the source-free Teukolsky equation for any $\psi_i$ will be a defining equation for a Hertz-like  potential labeled by $\Psi_H$. Accordingly, each $\psi_i$ could generate a Hertz potential. Wald proved that by means of applying linear PDE operators on this $\Psi_H$ all $\psi$'s will be defined. If the conjugate source-free Teukolsky equation for $\psi_4$ is chosen to define the corresponding conjugate Hertz-potential $\Psi^{*}_H$, then we have

\begin{equation}
\begin{gathered}
\mathcal{O}_{-2}^* \Psi_H^*=0, \\
\mathcal{O}_{-2}^* \equiv\left[\left(\delta+3 \alpha^*+\beta-\tau\right)\left(\delta^*+4 \beta^*+3 \tau^*\right)-\right. \\
\left.\left(\Delta-\gamma+3 \gamma^*+\mu\right)\left(D+4 \epsilon^*+3 \rho^*\right)+3 \psi_2^*\right\}.
\end{gathered}
\end{equation}

Then both $\psi_0$ and $\psi_4$ will be provided respectively as
\begin{equation}
\begin{gathered}
\psi_0= \\
\frac{1}{2}\left[\left(D-3 \epsilon+\epsilon^*-\rho^*\right)\left(D-2 \epsilon+2 \epsilon^*-\rho^*\right)\right. \\
\left.\left(D-\epsilon+3 \epsilon^*-\rho^*\right)\left(D+4 \epsilon^*+3 \rho^*\right)\right] \Psi_{H}^*.
\end{gathered}
\end{equation}

\begin{equation}
\begin{gathered}
\psi_4= \\
\frac{1}{2}\left[\left(\delta^*+3 \alpha+\beta^*-\tau^*\right)\left(\delta^*+2 \alpha+2 \beta^*-\tau^*\right)\right. \\
\left.\left(\delta^*+\alpha+3 \beta^*-\tau^*\right)\left(\delta^*+4 \beta^*+3 \tau^*\right)\right] \Psi_{H}^* \\
+3 \psi_2\left[\tau\left(\delta^*+4 \alpha\right)-\rho(\Delta+4 \gamma)-\mu(D+4 \epsilon)+\right. \\
\left.\pi(\delta+4 \beta)+2 \psi_2\right] \Psi_{H}.
\end{gathered}
\end{equation}
These equations are known as the ingoing radiation gauge (IRG) provided by Wald in \cite{wald_1973}, and are relating the gravitational Hertz potential $\Psi_{H}$ to Weyl scalars $\psi_i$ in the NP formalism. The tetrad legs are aligned along the repeated null direction of the Weyl tensor. The equations connecting $\Psi_H$ to $\psi_4$ will be the same angular equation that appears in BL coordinates
\begin{equation}
\begin{gathered}
(r-ia\cos\theta)^{4} \psi_4=\frac{1}{8}\left[\tilde{L}^4 \Psi^{*}_H-12 M \partial_t \Psi_H\right].\\
\end{gathered}
\end{equation}
where the $\tilde{L}^4$ is given by 
\begin{equation}
\begin{gathered}
\tilde{L}^4=L_{1} L_{0} L_{-1} L_{-2},\\
L_n \equiv -\partial_{\theta}+a\omega \sin \theta - \frac{m}{\sin \theta} +n \cot \theta.
\end{gathered}
\end{equation}
The  equation connecting $\Psi_H$ to $\psi_0$ will still maintain its radial nature but will take a different form as shown below. 
In the IEF coordinates,
\begin{equation}
\begin{gathered}
\left\{\frac{1}{2} D^4+\triangle^{\prime} D^3+\left[6 \triangle+2\left(a^2-m^2\right)\right] D^2\right. \\
\left.+\left[6 \triangle^{\prime} \triangle+4 \triangle^{\prime}\left(a^2-m^2\right)\right] D+12 \triangle^2\right\} \Psi_H=\psi_0, \\
D=(\triangle+4 M r) \partial_t+\triangle \partial_r+2 a \partial_\phi.
\end{gathered}
\end{equation}
In the OEF coordinates,
\begin{equation}
\begin{gathered}
    \frac{1}{2} D^4  \Psi_H= \psi_0,\\
    D=\partial_t-\partial_r.
\end{gathered}
\end{equation}
Since the angular equation is form-invariant under these transformations, it will be useful to choose the method used in \cite{pound_merlin_barack_2014} to algebrize the angular fourth-order ODE.

\subsection{CCKW}

Since the Hertz-angular equation is already form-invariant, its algebraization would be very similar \cite{pound_merlin_barack_2014}. We can follow the same steps to algebrize the Hertz-angular equation using the Teukolsky-Starobinsky identities.  At this point, we can use the identity equivalent to equation (59) in \cite{chandrasekhar_2009}. \begin{equation}
L_{1} L_{0} L_{-1} L_{-2} S_{-2}=D_2 S_{+2},
\end{equation}
where D is defined with
\begin{equation}
\begin{aligned}
D^2&=\lambda_{C H}^2\left(\lambda_{C H}+2\right)^2+8 a \omega(m-a \omega) \lambda_{C H}\left(5 \lambda_{C H}+6\right)\\
&+48 a^2 \omega^2\left[2 \lambda_{C H}+3(m-a \omega)^2\right],
\end{aligned}
\end{equation}
where $\lambda_{C H}=\lambda+s+2$. Also, we can write the Hertz potential as 
\begin{equation}
\begin{gathered}
\Psi^{\pm}_{H}=\sum_{lmw} H_{lmw} {}_{-2} \tilde{\tilde{R}}_{lmw} e^{i (m\phi-\omega t)}{ }_{-2} S_{lmw}(\theta).\\
\end{gathered}
\end{equation}
given that ${}_{-2} S_{lm\omega}^{*}(\theta)= (-1)^m { }_{2} S_{lm\omega}(\theta)$.
Then, the angular equation can be written as 
\begin{equation}
\begin{gathered}
\sum_{lm\omega} \{8 (r-ia\cos\theta)^4 \mathcal{C}_{lm\omega} {}_{-2} \mathcal{R}_{lm\omega}+12 iM\omega {}_{-2} \tilde{\tilde{R}}_{l m \omega} H_{lm\omega}\\
-(-1)^m D {}_{-2} \tilde{\tilde{R}}^{*}_{l -m -\omega} H_{l-m-\omega}^{*}
 \}e^{i (m\phi-\omega t)}{ }_{-2} S_{lm\omega}(\theta)=0.\\
\end{gathered}
\end{equation}
Then we arrive to this relation
\begin{equation}
\begin{gathered}
8 (r-ia\cos\theta)^4 \mathcal{C}_{lm\omega} {}_{-2} \mathcal{R}_{lm\omega}=-12 iM\omega {}_{-2} \tilde{\tilde{R}}_{l m \omega} H_{lm\omega}\\
+(-1)^m D {}_{-2} \tilde{\tilde{R}}^{*}_{l -m -\omega} H_{l-m-\omega}^{*}.
\end{gathered}
\end{equation}
We can take the complex conjugate of this equation and solve for $H_{lmw}  {}_{-2} \tilde{\tilde{R}}_{lm\omega} $, and finally write $\Psi^{\pm}_{H}$ as 
\begin{equation}
\begin{gathered}
\Psi^{\pm}_{H}=\sum_{lm} [\mathcal{A}_{lm} \hspace{.1cm} {}_{-2} \mathcal{R}_{lm} + \mathcal{B}_{lm} \hspace{.1cm} {}_{-2} \mathcal{R}_{lm}^{*}] e^{i (m\phi-\omega t)}{ }_{-2} S_{lm}(\theta),\\
\mathcal{A}_{lm}=\frac{-96imM\omega (r-ia \cos \theta)^4 \mathcal{C}_{lm}}{D^2 + 144 M^2 m^2 \omega^2},\\ 
\mathcal{B}_{lm}= (-1)^m\frac{8D (r+ia \cos \theta)^4 \mathcal{C}_{lm}^{*}}{D^2 + 144 M^2 m^2 \omega^2}.\\
\end{gathered}
\end{equation}
We can use the relation $R_{lmw}^{*}=R_{l-m -w}$ to rewrite the expression for $\Psi^{\pm}_{H}$ as 
\begin{equation}
\begin{gathered}
\Psi^{\pm}_{H}=\sum_{lm\omega} \hspace{.1cm} { }_2 \mathcal{S}_{lm\omega} \hspace{.1cm} {}_2 \mathcal{R}_{lm\omega}e^{i (m\phi-\omega t)},\\
{ }_2 \mathcal{S}_{lm\omega}=\mathcal{A}_{lm\omega} \hspace{.05cm} { }_2 S_{lm\omega}(\theta) + \mathcal{B}_{lm\omega} \hspace{.05cm} { }_2 S_{l-m-\omega}(\theta).
\end{gathered}
\end{equation}

\subsection{Metric Reconstruction}

In the outgoing Radiation gauge, the metric perturbation could be constructed from the Hertz potential $\Psi^{\pm}_{H}$ following the CCKW procedure with this relation 
\begin{equation}
h^{\mu \nu}= S^{\mu \nu \dagger}_{+2} \Psi^{\pm}_{H}+ c.c.
\end{equation}
We can use the radial and angular ODEs as well as Fourier decomposition to write 
\begin{equation}
\begin{gathered}
h^{\mu \nu}=\sum_{lm\omega} \{\alpha_{lm\omega}^{\mu \nu} \hspace{.05cm} {}_{-2} \mathcal{R}_{lm\omega} \hspace{.05cm} { }_{-2} \mathcal{S}_{lm\omega}+\gamma_{lm\omega}^{\mu \nu}  \hspace{.05cm} {}_{-2} \mathcal{R}_{lm}^\prime \hspace{.05cm} { }_{-2} \mathcal{S}_{lm\omega}\\
+\beta_{lm\omega}^{\mu \nu} \hspace{.05cm} {}_{-2} \mathcal{R}_{lm} \hspace{.05cm} { }_{-2} \mathcal{S}_{lm\omega}^{\prime}\}+c.c.
\end{gathered}
\end{equation}

Each of $\alpha_{lm\omega}^{\mu \nu}$, $\beta_{lm\omega}^{\mu \nu}$ and $\gamma_{lm\omega}^{\mu \nu}$ are functions depending on variables $(r,\theta)$ and parameters $(\omega,m)$. These functions have no singular points away from the horizon. The metric suffers from the discontinuity at $r=R$ as we expected. We see that the perturbation of the metric is written in terms of the radial function and its derivative which have an additional singular point. Thus, the metric expansion needs to be treated carefully taking into consideration this additional singularity.

\subsection{Completion}

Although $\psi_0$ and $\psi_4$ contain most of the information about the gravitational perturbation, there are still missing parts due to the perturbation of the background itself. The regular parts of these perturbations come from the perturbation of the mass $M$ and angular momentum $J=aM$ of the black hole. Accordingly, the full metric perturbation ${}^{Full} h^{\mu \nu}$ can be written in this form
\begin{equation}
{}^{Full} h^{\mu \nu}= h^{\mu \nu}+c_M h^{\mu \nu (\delta M)}+c_J h^{\mu \nu (\delta J)}.
\end{equation}
Thus, we need to compute these parts to have the full regular metric perturbation. The perturbation due to the mass $h_{\mu \nu}^{(\delta M)}$ and angular momentum $h_{\mu \nu}^{(\delta J)}$ are given respectively by \cite{merlin_ori_barack_pound_van_de_meent_2016}.
\begin{equation}
\begin{aligned}
h_{\mu \nu}^{(\delta M)} &=\left.\frac{\partial g_{\mu \nu}\left(x^\mu ; M, J\right)}{\partial M}\right|_{J \rightarrow 0},\\
h_{\mu \nu}^{(\delta J)}&=\left.\frac{\partial g_{\mu \nu}\left(x^\mu ; M, J\right)}{\partial J}\right|_{J \rightarrow 0}.
\end{aligned}
\end{equation}
\vspace{1cm}

In the IEF coordinates,
\begin{equation}
h_{\mu \nu}^{(\delta M)} =\left(\begin{array}{cccc}
-\frac{2 r}{\Sigma} & -\frac{2 r}{\Sigma} & 0 & 0 \\
-\frac{2 r}{\Sigma} & -\frac{2 r}{\Sigma} & 0 & 0 \\
0 & 0 & 0 & 0 \\
0 & 0 & 0 & 0
\end{array}\right).
\end{equation}

\begin{equation}
h_{\mu \nu}^{(\delta J)} =
\left(\begin{array}{cccc}
0 & 0 & 0 & \frac{2 M r \sin ^2(\theta)}{\Sigma} \\
0 & 0 & 0 & 0 \\
0 & 0 & 0 & 0 \\
\frac{2 M r \sin ^2(\theta)}{\Sigma} & 0 & 0 & 0
\end{array}\right).
\end{equation}
In the OEF coordinates,

\begin{equation}
h_{\mu \nu}^{(\delta M)} =\left(\begin{array}{cccc}
-\frac{2 r}{\Sigma} & \frac{2 r}{\Sigma} & 0 & 0 \\
\frac{2 r}{\Sigma} & -\frac{2 r}{\Sigma} & 0 & 0 \\
0 & 0 & 0 & 0 \\
0 & 0 & 0 & 0
\end{array}\right).
\end{equation}

\begin{equation}
h_{\mu \nu}^{(\delta J)} =
\left(\begin{array}{cccc}
0 & 0 & 0 & \frac{2 M r \sin ^2(\theta)}{\Sigma} \\
0 & 0 & 0 & 0 \\
0 & 0 & 0 & 0 \\
\frac{2 M r \sin ^2(\theta)}{\Sigma} & 0 & 0 & 0
\end{array}\right).
\end{equation}

The above expressions represents the full metric perturbation in the IEF coordinates up to undetermined coefficients $c_M$ and $c_J$. 

\section{Conclusion and Discussion}

In this work, we studied perturbations of the Kerr metric due to a circularly orbiting perturber in different spacetime foliations: Boyer–Lindquist and outgoing/ingoing Eddington-Finkelstein coordinates. This problem may have applications in many realistic astrophysical situations.  The reason for utilizing different foliations was to make a contrast between the regular and irregular charts and tetrad at the horizon. Though the Teukolsky equation has the same singularity structure, the asymptotic behavior of the equations at the horizon was different. We showed that, for the Kerr black hole perturbations in regular charts near the horizon, the radial part of the Weyl scalars naturally obeys the physical boundary conditions at the horizon and in the asymptotically flat regions. This removes the need for imposing any regularization conditions. Consequently, the freedom of a null rotation is still present which might be used as a gauge freedom. 
\vspace{1mm}

Using the CCKW procedure, we explicitly constructed the Kerr metric perturbation due to the existence of a perturber of energy $E$ rotating around the black hole in circular orbits. We effectively expanded the metric in all the Weyl-scalars perturbation modes with spacetime-dependent coefficients. In our construction of the metric, we used the Green's functions method as well as a Hertz-Weyl equation algebraization technique identical to the ones provided in \cite{shah_friedman_keidl_2012}. However, we solved for $\psi_{4}$ in a different manner as illustrated through equations (\ref{60}-\ref{65}). Moreover, we completed the metric by fixing the trivial physical perturbation due to the mass and angular momentum of the black hole itself, in a way similar to the work found in \cite{van_de_meent_shah_2015}. We didn't determine the two coefficients $c_M$ and $c_J$, which (if needed) could be evaluated using the procedure introduced in \cite{merlin_ori_barack_pound_van_de_meent_2016} by utilizing the gauge-invariant quantities. We also ignored the divergence contribution from the C-metric acceleration and NUT-charge for physical considerations.
\vspace{1mm}

 The radial equations in the IEF/OEF as well as BL coordinates are the confluent Heun equations. Consequently, if we are only interested in obtaining the Teuklosky equation, there exists a radial transformation that can transform the equation directly, based on the nature of the Heun family ODEs. We are not reporting this transformation here, yet we believe it is a straightforward though tedious exercise following the procedure given in \cite{ronveaux_2007}. The first derivative of the radial equation has an additional regular singular point whose location depends on the spacetime parameters, $(M,a)$, and the perturbation mode parameters, $(m,\omega)$. 
 
 To extend investigation of the perturbations beyond the horizon in the presented formalism, an explicit solution regular at the outer horizon might be needed. This can perhaps be achieved by a singular series expansion of the perturbed radial part of the equations, as well as its derivative. The existence of the derivative of the radial functions in the metric expansion is crucial for the radius of convergence of the expansion. Also, the expansion itself will be undefined at $r=R$ where the perturber orbit is located. We report, in that the procedure outlined here; an explicit form of metric construction as expansion in the solution for the radial functions, their derivative as well as the angular functions. 
\vspace{1mm}

\begin{acknowledgments}
We wish to thank Professor Gino Biondini for the useful discussions about the mathematical tools used in this paper. We are also grateful to Omar Elserif for helping us with proofreading. D.S. is partially supported by the US National Science Foundation, under Grant No.  PHY-2014021.  
\end{acknowledgments}

\bibliography{ref}

\end{document}